\documentclass[prl,aps,twocolumn,showpacs,a4paper]{revtex4}
\usepackage{bm}
\usepackage[dvips]{graphicx}
\usepackage{amsmath}

\begin{document}

\title{Dancing {\it Volvox}~: Hydrodynamic Bound States of Swimming Algae}

\author{Knut Drescher$^{1}$, Kyriacos C. Leptos$^{1}$, Idan Tuval$^{1}$, Takuji Ishikawa$^{2}$, 
Timothy J. Pedley$^{1}$, and Raymond E. Goldstein$^{1}$}
\affiliation{$^{1}$Department of Applied Mathematics and Theoretical
Physics, University of Cambridge, Cambridge CB3 0WA, UK}
\affiliation{$^{2}$Department of Bioengineering and Robotics, Tohoku University, Sendai 980-8579, Japan}

\date{\today}

\begin{abstract}
The spherical alga {\it Volvox} swims by means of flagella on 
thousands of surface somatic cells.  This geometry and its large size 
make it a model organism for studying the fluid dynamics of multicellularity.  
Remarkably, when two nearby {\it Volvox} swim close to a solid surface, 
they attract one another and can form stable bound states in which 
they ``waltz" or ``minuet" around each other.  
A surface-mediated hydrodynamic attraction
combined with lubrication forces between spinning, bottom-heavy 
{\it Volvox} explains the formation, stability and dynamics of the 
bound states.  These phenomena are suggested to underlie observed 
clustering of {\it Volvox} at surfaces.
\end{abstract}


\maketitle

\begin{figure}[b]
\begin{center}
\includegraphics*[clip=true,width=0.95\columnwidth]{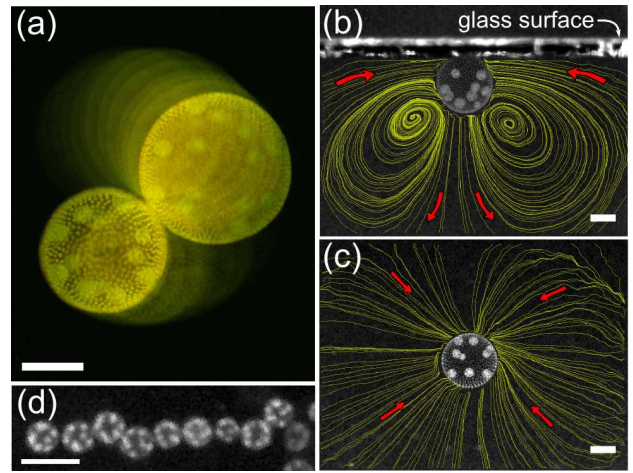}
\end{center}
\caption{\label{fig1} (Color online) Waltzing of {\it V. carteri}.  
(a) Top view. Superimposed images
taken $4$ s apart, graded in intensity. (b) Side, and (c) top views of 
a colony swimming against a coverslip, with 
fluid streamlines.  Scales are $200$ $\mu$m.  (d) A linear {\it Volvox} cluster 
viewed from above (scale is $1$ mm).}
\end{figure}

Long after he made his great contributions to microscopy and started a revolution in 
biology, Antony van Leeuwenhoek peered into a drop of pond water and discovered one of 
nature's geometrical marvels \cite{Leeuwenhoek}.  This was the freshwater alga which, 
years later, in the very last entry of his great work on biological taxonomy, 
Linneaus named {\it Volvox} \cite{Linneaus} for its characteristic spinning motion 
about a fixed body axis. {\it Volvox} is a spherical colonial green alga 
(Fig. \ref{fig1}), with thousands of biflagellated cells anchored 
in a transparent extracellular matrix (ECM) and daughter colonies inside the 
ECM. Since the work of Weismann \cite{Weismann}, {\it Volvox} has been seen as 
a model organism in the study of the evolution of 
multicellularity \cite{Kirkbook,twelvestep,multicellular}.

Because it is spherical, {\it Volvox} is an ideal organism for studies of 
biological fluid dynamics, being an approximate realization of Lighthill's 
``squirmer" model \cite{Lighthill} of self-propelled bodies having a specified surface 
velocity.  Such models have elucidated nutrient uptake at high P{\'e}clet 
numbers \cite{Magar,multicellular} by single organisms, and pairwise hydrodynamic 
interactions between them \cite{Ishikawa_pairwise}. Volvocine algae may also 
be used to study {\it collective} dynamics of self-propelled objects 
\cite{IshikawaPedley08}, complementary to bacterial suspensions ({\it E. coli, 
B. subtilis}) exhibiting large-scale coherence in thin films 
\cite{Wu} and bulk \cite{Dombrowski}.

While investigating {\it Volvox} suspensions in glass-topped chambers we observed 
stable bound states, in which pairs of colonies orbit each other near the 
chamber walls. {\it Volvox} is ``bottom-heavy" due to clustering of daughter 
colonies in the posterior, so an isolated colony swims upward with its axis 
vertical, rotating clockwise (viewed from above) at an angular frequency 
$\omega \sim 1$ rad/s for a radius $R\sim 150$ $\mu$m.  When approaching the 
chamber ceiling, two {\it Volvox} are drawn together, nearly touching while 
spinning, and they ``waltz" about each other clockwise (Fig. \ref{fig1}a) at 
an angular frequency $\Omega\sim 0.1$ rad/s. When {\it Volvox} have become 
too heavy to maintain upswimming, two colonies hover above one another near the 
chamber bottom, oscillating laterally out of phase in a ``minuet'' dance. Although 
the orbiting component of the waltzing is reminiscent of vortex pairs in 
inviscid fluids, the attraction and the minuet are not, and as the 
Reynolds number is $\sim 0.03$, inertia is negligible.  

While one might imagine that signalling and chemotaxis could result in these 
bound states, a combination of experiment, theory, and numerical computations 
is used here to show that they arise instead from the interplay of short-range 
lubrication forces between spinning colonies and surface-mediated hydrodynamic 
interactions \cite{Blake}, known to be important for colloidal 
particles \cite{KehAnderson,dufresne} and bacteria \cite{lauga_prl}. We 
conjecture that flows driving {\it Volvox} clustering at surfaces enhance 
the probability of fertilization during the sexual phase of their life cycle.

\begin{figure}[t]
\begin{center}
\includegraphics*[clip=true,width=0.95\columnwidth]{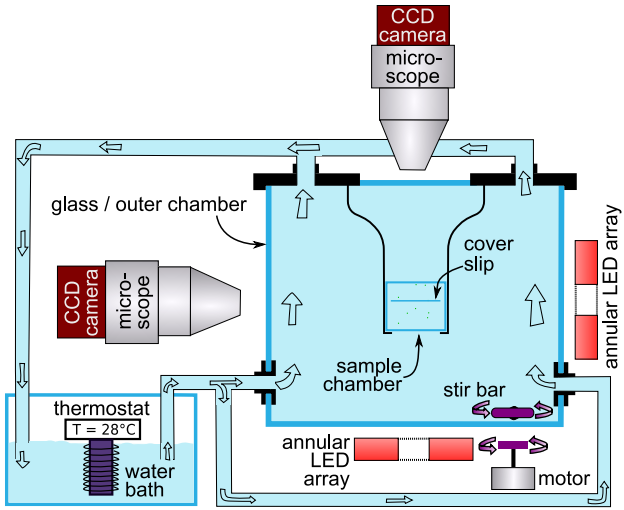}
\end{center}
\caption{\label{fig2} (Color online) Dual-view apparatus.}
\end{figure}

{\it Volvox carteri} f. {\it nagariensis} EVE strain (a subclone of HK10) were 
grown axenically in SVM \cite{kirk83,multicellular} in diurnal growth chambers 
with sterile air bubbling, in a daily cycle of $16$ h in cool 
white light ($\sim 4000$ lux) at $28^{\circ}$ C and $8$ h in the dark at $26^{\circ}$ C. 
Swimming was studied in a dual-view system (Fig. \ref{fig2}) \cite{RevSciInst}, 
consisting of two identical assemblies, each a CCD camera (Pike 145B, Allied 
Vision Technologies, Germany) and a long-working distance microscope (InfiniVar CMS-2/S, 
Infinity Photo-Optical, Colorado). Dark-field illumination used $102$ mm diameter 
circular LED arrays (LFR-100-R, CCS Inc., Kyoto) with narrow bandwidth emission at 
$655$ nm, to which {\it Volvox} is insensitive \cite{photospectrum}.  Thermal 
convection induced by the illumination was minimized by placing the 
$2\times 2\times 2$ cm sample chamber, made from microscope slides held together 
with UV-curing glue (Norland), within a stirred, temperature-controlled water 
bath. A glass cover slip glued into the chamber provided a clean surface (Fig. 
\ref{fig1}b) to induce bound states. Particle imaging velocimetry (PIV) studies 
(Dantec Dynamics, Skovelund, Denmark) showed that the r.m.s convective velocity 
within the sample chamber was $\lesssim 5$ $\mu$m/s.  

Four aspects of {\it Volvox} 
swimming are important in the formation of bound states, each arising, in the 
far field, from a distinct singularity of Stokes flow:
(i) negative buoyancy (Stokeslet), (ii) self-propulsion (stresslet), 
(iii) bottom-heaviness (rotlet), and spinning (rotlet doublet). During the 
$48$ hour life cycle, the number of somatic cells is constant; only their 
spacing increases as new ECM is added to increase the colony radius. This slowly 
changes the speeds of sinking, swimming, self-righting, and spinning, allowing 
exploration of a range of behaviors. The upswimming velocity $U$ was measured 
with side views in the dual-view apparatus. {\it Volvox} density was determined 
by arresting self-propulsion through transient deflagellation with a pH 
shock \cite{multicellular,Witman}, and measuring sedimentation.
%in a $4.7$ mm diameter cylindrical 
%polypropylene chamber, with one face milled off and replaced by a coverslip. 
The settling velocity $V=2\Delta\rho g R^2/9\eta$, with $g$ the
acceleration of gravity and $\eta$ the fluid viscosity, yields the
density offset $\Delta\rho=\rho_c-\rho$ between the colony and water.
Bottom-heaviness implies a distance $\ell$ between the 
centers of gravity and geometry,
measured by allowing {\it Volvox} to roll off
a guide in the chamber and monitoring the 
axis inclination angle $\theta$ with the vertical. This angle obeys
$\zeta_r\dot \theta=-(4\pi R^3\rho_c g\ell/3)\sin\theta$, where $\zeta_r=8\pi \eta 
R^3$ is the rotational drag coefficient, leading to a relaxation time 
$\tau=6\eta/\rho_c g\ell$ \cite{gyrotaxis}. 
The rotational frequencies $\omega_o$ of free-swimming colonies 
were obtained from movies, using germ cells/daughter colonies as markers. 

\begin{figure}[t]
\begin{center}
\includegraphics*[clip=true,width=1.00\columnwidth]{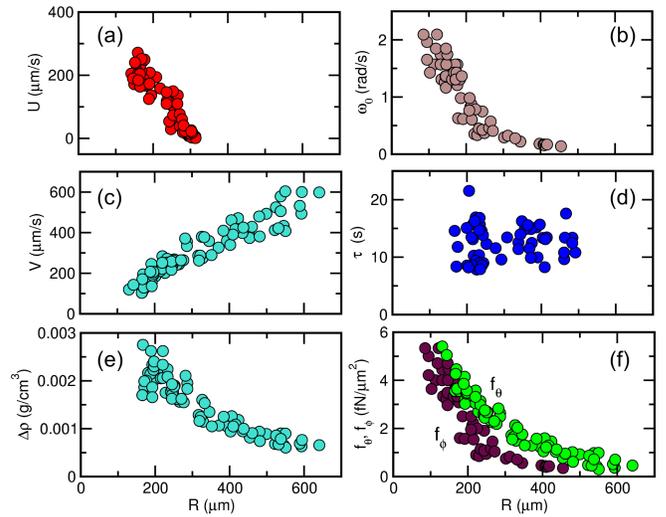}
\end{center}
\caption{\label{fig3} (Color online) Swimming properties of {\it V. carteri} 
as a function of radius. 
(a) upswimming speed, (b) rotational frequency, 
(c) sedimentation speed, (d) reorientation time, 
(e) density offset, and (f) components of average 
flagellar force density.}
\end{figure}

Figure \ref{fig3} shows the four measured quantities
($U,V,\omega_o,\tau$) and the deduced density offset 
$\Delta \rho$. 
In the simplest model \cite{multicellular},
locomotion derives from a uniform force per unit area ${\bf f}=
(f_{\theta},f_{\phi})$ exerted by flagella tangential to the colony surface.  
Balancing the net force $\int dS\,{\bf f}\cdot \hat{\bf z}
=\pi^2 f_{\theta}R^2$ against the Stokes drag and negative buoyancy yields 
$f_{\theta}=6\eta (U+V)/\pi R$.
Balancing the flagellar torque 
$\int dS\, R ( \hat{\bf r} \times {\bf f} ) \cdot \hat{\bf z}=\pi^2 f_{\phi}R^3$
against viscous rotational torque 
$8\pi\eta R^3\omega_o$ yields $f_{\phi}=8\eta\omega_o/\pi$. 
These components are shown in Fig. \ref{fig3}f, where we used a
linear parameterization of the upswimming data (Fig. \ref{fig3}a) 
to obtain an estimate of $U$ over the entire radius range.
The typical force density $f_{\theta}$ 
corresponds to several pN per flagellar pair \cite{multicellular}, while the
relative smallness of $f_{\phi}$ is a consequence of
the $\sim15^\circ$ tilt of the flagellar beating plane with respect to the 
colonial axis \cite{Hoops,jfm}. 

\begin{figure}[t]
\begin{center}
\includegraphics*[clip=true,width=0.9\columnwidth]{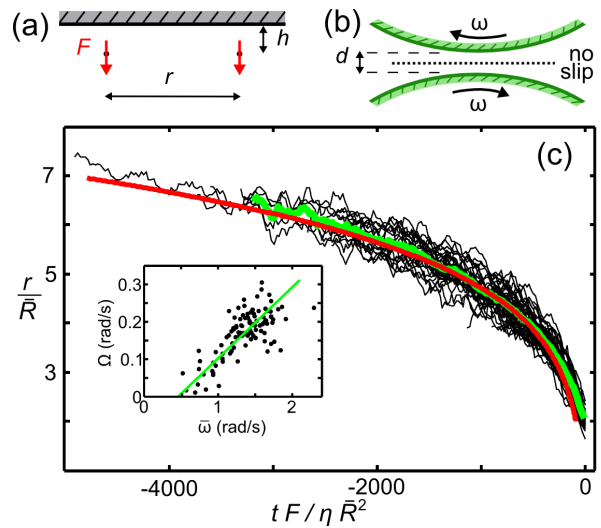}
\end{center}
\caption{\label{fig4} (Color online) Waltzing dynamics. Geometry of (a) two 
interacting Stokeslets (side view) and (b) nearby spinning 
colonies. (c) Radial separation $r$, normalized by mean colony radius, as a 
function of rescaled time for $60$ events (black). Running average (green) 
compares well with predictions of the singularity model (red). Inset shows orbiting 
frequency $\Omega$ as a function of mean spinning frequency $\bar\omega$, and linear fit.}
\end{figure}

Using the measured parameters it is possible to characterize both bound states. 
Fig. \ref{fig4}c shows data from measured tracks of $60$ pairs of {\it Volvox}, 
as they fall together to form the waltzing bound state. The data collapse when 
the inter-colony  separation $r$, normalized by $\bar R$, the mean of the two 
participating colonies' radii, is plotted as a function of rescaled time from 
contact.  The waltzing frequency $\Omega$ is linear in the mean spinning 
frequency of the pair $\bar \omega$. These two ingredients of the waltzing bound 
state, ``infalling'' and orbiting, can be understood, respectively, by 
far-field features of mutually-advected singularities and near-field effects given 
by lubrication theory, which will now be considered in turn.

{\it Infalling:} When swimming against an upper surface, the net thrust induced by 
the flagellar beating is not balanced by the viscous drag on the colony, 
as the colony is at rest, resulting in a net downwards force on the fluid. The 
fluid response to such a force may be modeled as a Stokeslet normal to and 
at a distance $h$ from a no-slip surface \cite{Blake}, forcing fluid in 
along the surface (Fig. \ref{fig1}c) and out below the colony, with a 
toroidal recirculation.  Seen in cross section with PIV, the velocity field 
of a single colony has precisely this appearance (Fig. \ref{fig1}b).  
This flow produces the attractive interaction between colonies; Squires has proposed 
a similar scenario in the context of electrophoretic levitation \cite{Squires}.

The motion of a pointlike object at ${\bf x}_i$, with axis orientation 
${\bf p}_i$ and net velocity ${\bf v}_i$ from self-propulsion and buoyancy, 
due to the fluid velocity ${\bf u}$ and vorticity ${\bf \nabla}\times {\bf u}$ 
generated by the other self-propelled objects, obeys
\begin{eqnarray}
\dot {\bf x}_i &=& {\bf u}({\bf x}_i)+{\bf v}_i ~ , \label{eom1} \\
\dot {\bf p}_i &=& {1\over \tau}
{\bf p}_i\times\left(\hat{\bf z} \times{\bf p}_i\right)
+{1\over 2}\left({\bf \nabla}\times{\bf u}\right)\times {\bf p}_i~. \nonumber
\end{eqnarray}
Assuming that for the infalling, ${\bf v}_i = \dot{{\bf p}}_i= 0$, and 
that ${\bf u}({\bf x}_i)$ are due to Stokeslets of strength 
$F= 6 \pi \eta R (U + V)$, Eq. \ref{eom1} may be reduced, in 
rescaled coordinates 
$\tilde r = r / h$ and $\tilde t = t F / \eta {h}^2$ with $h= \bar R$, to \cite{Squires} 
\begin{equation}
\frac{\mbox{d} \tilde r}{\mbox{d} \tilde t} = -\frac{3}{\pi} \frac{ \tilde r}{({\tilde r}^2 + 4)^{5/2}}~.
\label{eom2}
\end{equation}
Integration of (\ref{eom2}) shows good parameter-free agreement with the 
experimental trajectories of nearby pairs (Fig. \ref{fig4}c). Large 
perturbations to a waltzing pair by a third nearby colony can disrupt it by strongly 
tilting the colony axes, suggesting that bottom-heaviness confers stability. 
This is confirmed by a linear stability analysis \cite{jfm}.  

\begin{figure*}[t]
\begin{center}
\includegraphics*[clip=true,width=1.90\columnwidth]{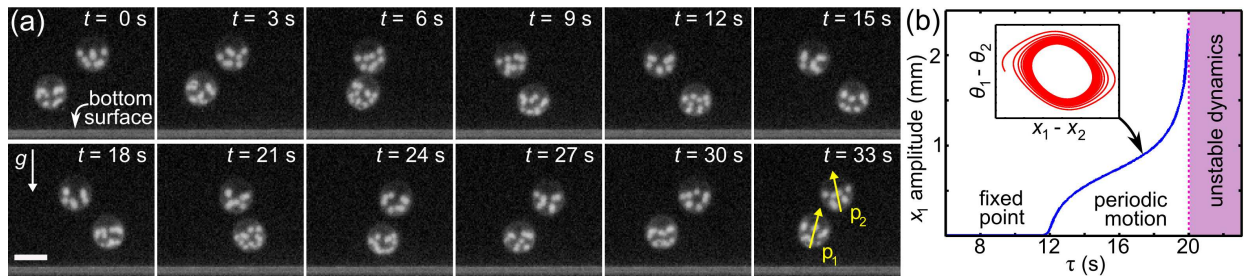}
\end{center}
\caption{\label{fig5} (Color online) ``Minuet" bound state.  (a) 
Side views $3$ s apart of two colonies near the chamber bottom.  Yellow arrows indicate the 
anterior-posterior axes ${\bf p}_i$ at angles $\theta_i$ to vertical. 
Scale bar is $600$ $\mu$m.  (b) Bifurcation diagram, and phase portrait (inset), 
showing a limit cycle, with realistic model parameters $F= 6 \pi \eta R V$, 
$R = 300$ $\mu$m, $h_1 = 450$ $\mu$m, $h_2 = 1050$ $\mu$m.}
\end{figure*}

{\it Orbiting:} As {\it Volvox} colonies move together under the influence of 
the wall-induced attractive flows (Fig. \ref{fig1}b), orbiting becomes noticeable 
only when their separation $d$ is $\lesssim 30$ $\mu$m; their spinning frequencies also 
decrease very strongly with decreasing separation. This arises from viscous torques 
associated with the thin fluid layer between two colonies (Fig. \ref{fig4}b). 
We assume that in the thin fluid layer, the spinning {\it Volvox} colonies can be 
modeled as rigid spheres, ignoring the details of the overlapping flagella 
layers. For two identical colonies, ignoring the anterior-posterior 
``downwash," and considering only the region where the fluid layer is thin, 
the plane perpendicular to the line connecting their centers is a locus of 
zero velocity, as with a no-slip wall. Appealing to known asymptotic results 
\cite{lube} we obtain the torque ${\cal T}=-(2/5)\ln(d/2R)\zeta_r\omega$ and 
a lateral force ${\cal F}=(1/10)\ln(d/2R)\zeta_r\omega/R$ on the sphere, 
where $\omega<\omega_o$ is the spinning frequency of a colony in the 
bound state.  The rotational slowing of the self-propelled colony has an effect on the 
fluid that may be approximated by a rotlet of strength ${\cal T}$ at 
its center. From the flow field of a rotlet perpendicular to a horizontal 
no-slip wall \cite{Blake} and the lateral force ${\cal F}$, we then 
deduce the orbiting frequency
\begin{equation}
\Omega\simeq 0.069\,\ln\left({d\over 2R}\right)\, \bar\omega~.
\label{torque}
\end{equation}
Typical values of $d$ and $R$ give a slope of $\simeq 0.14-0.19$ for 
the $\Omega-\omega$ line, consistent with the experimental fit of $0.19 \pm 0.05$ 
(Fig. \ref{fig4}c). The nonzero intercept is likely due to lubrication friction
against the ceiling \cite{jfm}.

A second and more complex type of bound state, the ``minuet," is found when 
the upswimming just balances the settling (at $R \simeq 300$ $\mu$m, 
see Fig \ref{fig3}a), and {\it Volvox} colonies hover at a fixed distance above 
the chamber bottom. In this mode (Fig. \ref{fig5}) colonies stacked one above 
the other oscillate back and forth about a vertical axis.  The mechanism of 
oscillation is the instability of the perfectly aligned state due to the 
vorticity from one colony rotating the other, whose swimming brings it back, 
with the restoring torques from bottom-heaviness conferring stability.  Studies of 
the coupled dynamics of $\mathbf{x}_i$ and $\mathbf{p}_i$ show that when the 
orientational relaxation time $\tau$ is below a threshold the stacked arrangement 
is stable, while for $\tau$ larger there is a Hopf bifurcation to limit-cycle 
dynamics (Fig. \ref{fig5}b). In these studies, lubrication effects were 
ignored, $\dot{\mathbf{x}}_i$ was restricted to be in one horizontal dimension 
only, and $\mathbf{x}_i$ were at fixed heights $h_i$ above the wall. The 
flow $\mathbf{u}$ was taken to be due to vertically oriented Stokeslets at 
$\mathbf{x}_i$, of magnitude $F$, equal to the gravitational force on the {\it Volvox}.

Hydrodynamic bound states, such as those described here, may have biological 
significance. When environmental conditions deteriorate, {\it Volvox} colonies 
enter a sexual phase of spore production to overwinter. Field studies show that 
bulk {\it Volvox} concentrations $n$ are $<1$ cm$^{-3}$ \cite{cornell_thesis}, 
with male/female ratio of $\sim 1/10$, and $\sim 100$ sperm packets/male. 
Under these conditions, the mean encounter time for females and sperm packets 
is a substantial fraction of the life cycle. The kinetic theory mean 
free path $\lambda=1/ \sqrt{2} n \pi (R+ R_{sp})^2 \times 10/100$, with 
$R = 150$ $\mu$m for females, and $R_{sp} = 15$ $\mu$m for sperm 
packets, is $\lambda \sim 1$ m, implying a mean encounter time $>2$ h \cite{tobias}. This 
suggests that another mechanism for fertilization must be at work, with previous 
studies having excluded chemoattraction in this system \cite{coggin}. At naturally 
occuring concentrations, more than one {\it Volvox} may partake in the 
waltzing bound state, leading to long linear arrays (Fig. \ref{fig1}d). In such 
clusters, formed at the air-water interface, the recirculating flows would decrease 
the encounter times to seconds or minutes, clearly increasing the chance of sperm 
packets finding their target. Studies are underway to examine this possibility.
%, an estimate substantially unchanged by more detailed aspects of swimming or fluid flow \cite{LewisPedley}.  

We thank D. Vella, S. Alben and C.A. Solari for key observations, 
A.M. Nedelcu for algae, and support from the BBSRC, DOE, and the 
Schlumberger Chair Fund.

\end{document}